%% file: main.tex
\documentclass[conference]{IEEEtran}
\IEEEoverridecommandlockouts

\input{packages.tex}

\begin{document}

\title{\textit{ARIM-mdx Data System}: Towards a Nationwide Data Platform for Materials Science}

\input{authorieee.tex}

\maketitle

\begin{abstract}
In modern materials science, effective and high-volume data management across leading-edge experimental facilities and world-class supercomputers is indispensable for cutting-edge research.
However, existing integrated systems that handle data from these resources have primarily focused just on smaller-scale cross-institutional or single-domain operations.
As a result, they often lack the scalability, efficiency, agility, and interdisciplinarity, needed for handling substantial volumes of data from various researchers.

In this paper, we introduce \textit{ARIM-mdx data system}\footnote{The service is available at \url{https://arim.mdx.jp}.}, aiming at a nationwide data platform for materials science in Japan.
Currently in its trial phase, the platform has been involving 11 universities and institutes all over Japan, and it is utilized by over 800 researchers from around 140 organizations in academia and industry, being intended to gradually expand its reach.
The ARIM-mdx data system, as a pioneering nationwide data platform, has the potential to contribute to the creation of new research communities and accelerate innovations.
\end{abstract}

\begin{IEEEkeywords}
Big Data, Materials Science, Supercomputer, Cloud Computing, Shared-use Research Facility
\end{IEEEkeywords}

\section{Introduction}
In this era of Machine Learning (ML) and Artificial Intelligence (AI) technologies, the importance of scientific data in materials science has grown significantly.
The remarkable success of ML and AI in fields such as image processing and natural language processing is deeply rooted in the availability of vast amounts of data~\cite{kaplan2020scaling}.
Consequently, there has been a significant surge in the development of comprehensive data systems that facilitate the collection, management, and analysis of materials science data at scale.

\begin{figure}
  \centering
   \includegraphics[width=.9\columnwidth]{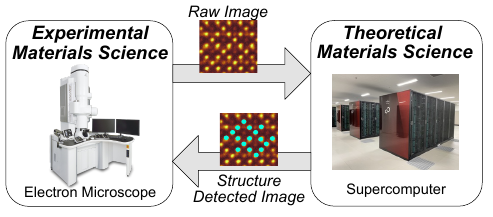}
  \vspace*{\vspacebeforecaptionsmall}
  \caption{Collaboration between experimental and theoretical materials science. The example shows a typical nanostructure analysis, where the atomic-scale image of the targeting material is obtained from the electron microscope whereas it is analyzed via computation-intensive theoretical simulation using the supercomputer to detect its structure. The raw image is obtained from~\cite{ishikawa2022direct}.}
  \label{fig:intro}
  \vspace*{\vspaceaftercaption}
\end{figure}

In materials science, the synergistic collaboration between theoretical and experimental methods is essential, as shown in Fig.~\ref{fig:intro}. 
Theoretical data, typically derived from supercomputer simulations, are obtained through computationally intensive calculations such as density functional theory (DFT) or molecular dynamics (MD). 
These calculations enable the prediction of material structures and various properties, including stability, formation energy, thermal conductivity, and magnetism. 
Conversely, real experimental data are obtained from various specialized equipment, such as electron microscopes and X-ray analytical instruments, among others.

The rapid advancement and increasing sophistication of modern research methods in materials science have made it crucial to utilize large-scale \textit{shared-use} resources, such as state-of-the-art experimental facilities and high-performance supercomputers. 
To enhance scientific research, it is vital not only to provide them as common tools but also to effectively collect, manage, analyze, and leverage the vast amounts of data generated through these advanced research methods.

Traditionally, large-scale data collection across a wide range of fields in materials science have been pursued predominantly for the computational simulation data~\cite{jain2013commentary,CURTAROLO2012218,kirklin2015open,qu2015electrolyte,draxl2019nomad,choudhary2020joint,merchant2023scaling}.
Moreover, nationwide platforms to accelerate such computational science research have been well-established in the high-performance computing fields~\cite{towns2014xsede,jhpcn,access}.
Considering the rapid advancements in computational methodologies, there arises an undeniable necessity to integrate this vast amount of the computational simulation data with real experimental data.

However, despite the substantial success in the wide collection and large-scale management of the computational data, that of the real experimental data remains a significant challenge. 
Most efforts on the collection, management and analysis of the experimental data are primarily focused just on the smaller-scale cross-institution~\cite{salim2019balsam,enders2020cross,antypas2021enabling,bear2021share,veseli2023streaming,al2023cross,brace2023linking,da2024frontiers}, single domain~\cite{iudin2016empiar,melo2023fostering,iudin2023empiar,madany2020neurokube,babu2023deep,kalinin2023machine}, or data curated from research articles~\cite{icsd,gravzulis2012crystallography}.
This leaves a huge potential for leveraging a large amount of the raw, unprocessed data untouched.

To address these gaps, we have been developing \textit{ARIM-mdx data system}, aiming at a nationwide materials-science data infrastructure capable of effectively handling both computational and experimental data.
The ARIM-mdx data system is designed to meet essential system requirements for data storage, management, and analysis.
The system requirements and our contributions towards this nationwide deployment are summarized as follows:

\smallskip
\noindent{\textbf{Stability and Security:}} 
The system must maintain high stability and security to manage a wide range of sensitivity levels in materials science data, which, for example, may occasionally include confidential information about hazardous substances or economically significant new materials. 

For the stability and security, the ARIM-mdx data system is based on a peta-scale dedicated storage system, rather than on public clouds or shared-use storage systems.
Such an isolated setup enables us to make physical operations be fully under control.
Moreover, to ensure safe data transfer from network-disabled facility computers, we offload the data transfer mechanism to IoT devices, thereby mitigating security risks on the facility computers.

\smallskip
\noindent{\textbf{Scalability and Efficiency:}}
Due to the rapid advancement of experimental facilities and supercomputers, these generate richer data at a higher rate.
For example, the state-of-the-art Cryogenic Electron Microscopy (Cryo-EM) generates up to TB-scale datasets per single material~\cite{iudin2016empiar,iudin2023empiar}.
The scalability and efficiency to process the large volume of data are clearly important for the nationwide service deployment.

In order to make the computing resources scalable while keeping the aforementioned isolated setups in the storage side, the ARIM-mdx data system adopts a hybrid architecture that separates computational and storage resources.
By allocating the former to a general-use academic cloud, the ARIM-mdx data system maintains scalable and efficient computation while ensuring the stability and security in the storage side.

\smallskip
\noindent{\textbf{Agility and Interdisciplinarity:}} 
As materials research entails collaboration across researchers from multiple domains in different institutions, the demands on the computational resources can fluctuate. 
This necessitates not only resource-intensive, predictable batch workload (i.e., traditional supercomputer workload), but also immediate resources for interactive data analysis (e.g., interactive data analysis during experiments).

The ARIM-mdx data system handles this mixed workload by providing the high-performance interactive computational resource from the cloud system while establishing high-performance network connection from supercomputers to efficiently transfer the results.

The remainder of this paper is summarized as follows.
\S\ref{sec:background} illustrates the background.
\S\ref{sec:overview} discusses the overview of the system.
In \S\ref{sec:infra}, the implementation of the system is described.
In \S\ref{sec:eval} and~\ref{sec:usecase}, we show the basic performance result and several use cases.
Finally, in \S\ref{sec:future},~\ref{sec:relatedwork} and~\ref{sec:conclusion}, we summarize our key findings, the suggested future directions and related work, before concluding this paper.

\section{Background: MEXT ARIM Japan Project}\label{sec:background}
Advanced Research Infrastructure for Materials and Nanotechnology in Japan (ARIM Japan 2021--2031)~\cite{ARIM2023} by the Ministry of Education, Culture, Sports, Science, and Technology (MEXT) is a continuation of a series of long-term initiatives (Nanotechnology Network Project in 2007--2011, Low Carbon Research Network Project in 2011--2016, Nanotech Technology Platform Project in 2012--2021), that aim to effectively manage shared-use experimental facilities located broadly in 13 universities and research institutes.
Over 1,000 units of experimental equipment are registered for shared use, and more than 2,000 research projects utilize them annually.

\section{ARIM-mdx Data System: Overview} \label{sec:overview}
\subsection{Key Design Principles}
\subsubsection{Usability}
To provide a common, large-scale, high-performance data infrastructure for both experimental and theoretical materials researchers, it is crucial to maximize usability while maintaining performance capable of handling large volumes of data.
Traditional tools commonly used on supercomputers are familiar to researchers in theoretical fields, but often less so to those in experimental domains. 
Therefore, constructing services based on user-friendly interfaces, such as web-based interfaces and GUI, like~\cite{hudak2018open}, is appropriate.

\subsubsection{Mixture of Interactive and Batch Workload}
In the typical research workflow in materials science, the interactive and batch computational workloads are mixed.
In the experimental research, for example, the immediate data analysis is performed during experiment using equipment (e.g., electron microscope), whereas in the theoretical research, the computation-intensive physical simulations (e.g., DFT or MD calculation) are performed on supercomputers.
Therefore, to effectively handle the data from both sources, it is crucial to provide flexible computational resources for both needs.

\subsubsection{Secure Data Transfer}\label{sec:secure}
Typical experimental equipment is managed under standalone setups because the operation computers are specifically tailored and highly optimized for particular setups, where, e.g., just a minor security OS update can harm the equipment control~\cite{hanai2023cloud}.
The direct remote data acquisition from such non-networked equipment is necessary to efficiently and securely transfer large-scale data to the ARIM-mdx data system.

\subsection{System Overview}
\begin{figure}
  \centering
   \includegraphics[width=\columnwidth]{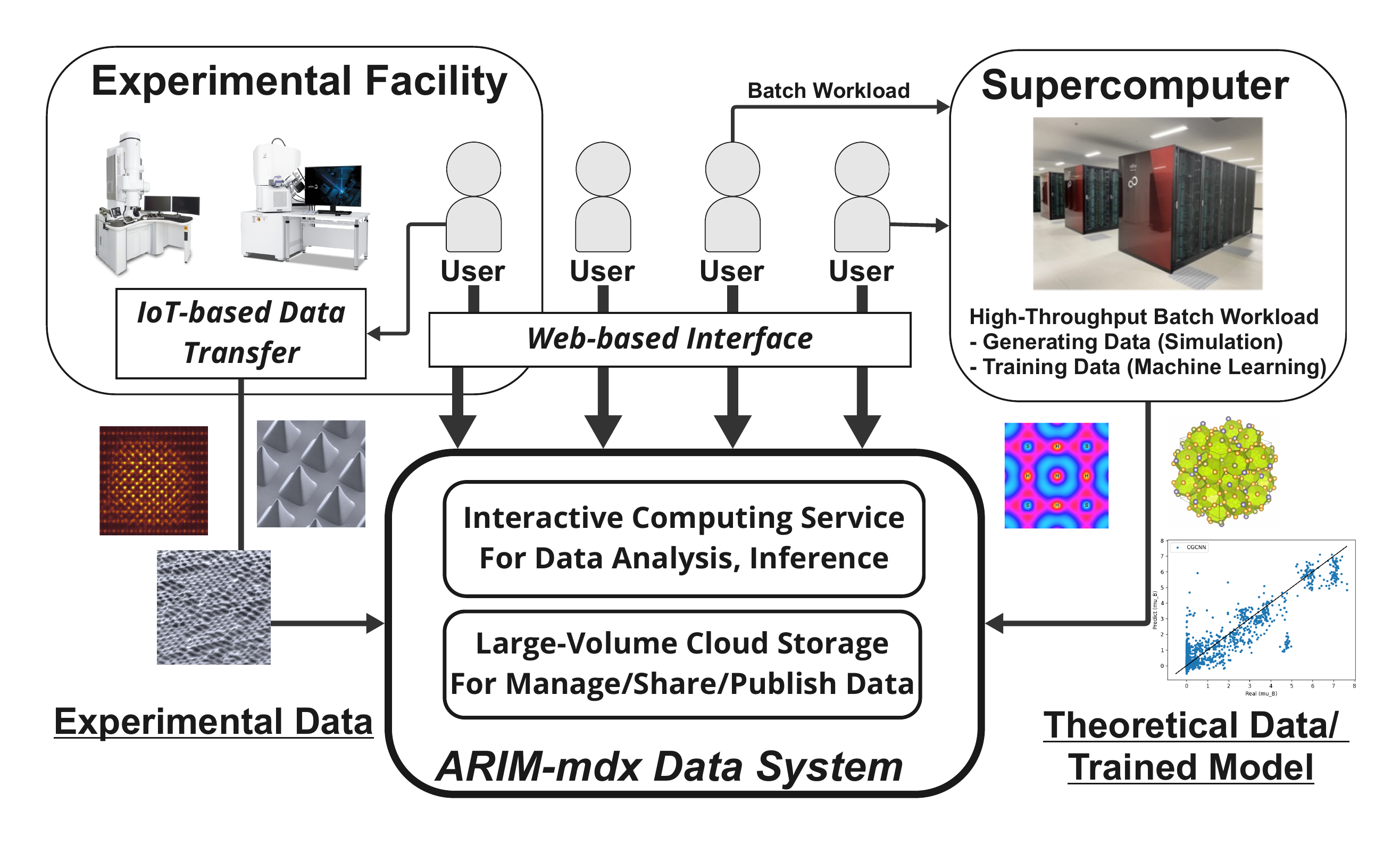}
  \vspace*{\vspacebeforecaption}
  \caption{System overview.}
  \label{fig:overview}
  \vspace*{\vspaceaftercaption}
\end{figure}

Fig.~\ref{fig:overview} shows the system overview.
The ARIM-mdx data system provides (i) interactive computing services, (ii) large-volume cloud storage, (iii) direct data transfer from experimental equipment, and (iv) high-throughput networking from the supercomputers in Japan.
The interactive computing services and cloud storage are provided via web-based interfaces.
The specifications of the interactive computation services can be flexibly configured for mid- to high-performance environments, including multiple GPUs or many cores.
Direct data transfer is enabled via data-transfer IoT devices~\cite{hanai2023cloud}, which provide an offloading method to remotely transfer data from non-networked standalone computers as well as per-user data distribution with user authentication.

The system is designed for three primary roles.

\subsubsection{As a research data storage and analysis service}
The system can be employed for typical data collection and analysis tasks, allowing researchers to transfer, store, and perform basic data analysis effectively.

\subsubsection{For data analysis within experimental facilities}
Researchers can utilize the data-transfer IoT devices to transfer data directly to the ARIM-mdx data system, enabling interactive data processing and analysis at the point of acquisition.

\subsubsection{As a service for managing and analyzing supercomputer-generated data}
The system supports high-speed data transfer from most supercomputers in Japan through SINET6, facilitating the efficient data management and analysis of large-scale computational results.
The efficient interactive analysis, typically limited on supercomputers, is a key significant function of the ARIM-mdx data system.

\begin{figure*}
  \centering
   \includegraphics[width=1.8\columnwidth]{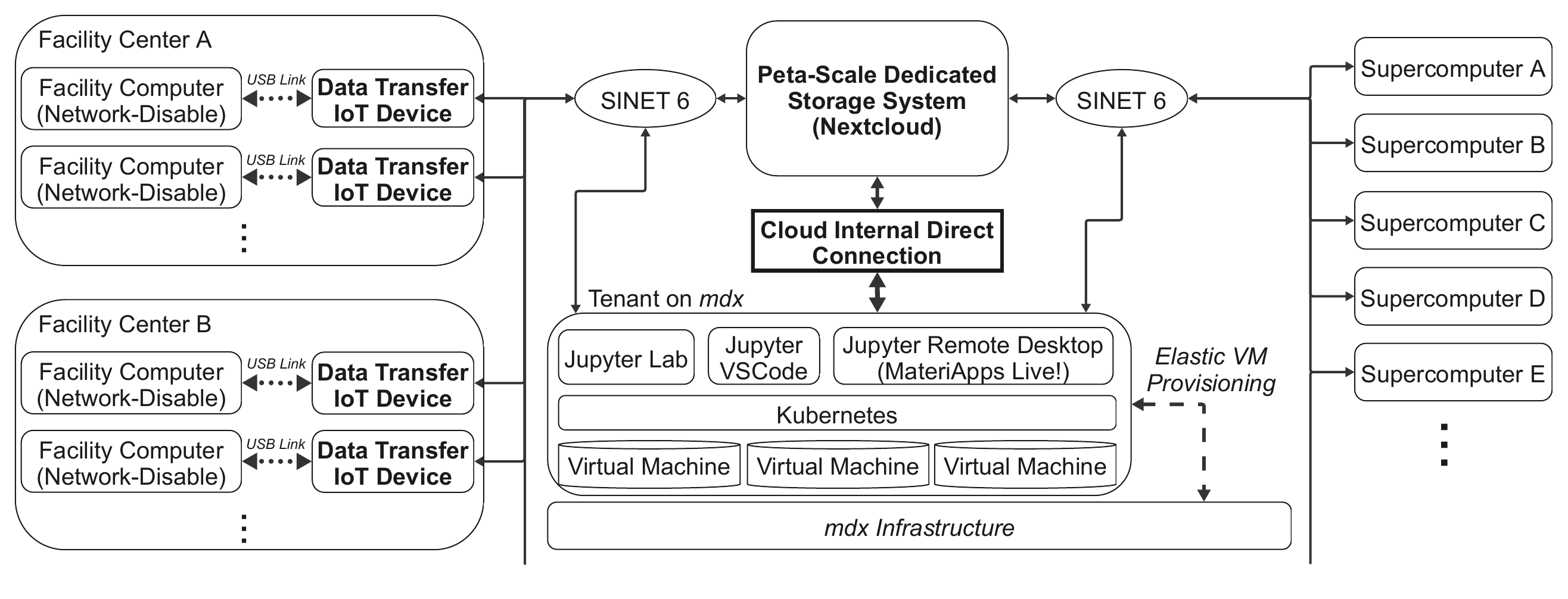}
  \vspace*{\vspacebeforecaption}
  \caption{System implementation.}
  \label{fig:implementation}
  \vspace*{\vspaceaftercaption}
\end{figure*}

\subsection{Universities and Institutions Involved} 
The ARIM-mdx data system has been involving several universities and institutes which operate research facilities and/or supercomputers as follows:
(1)~the University of Tokyo, (2)~SPring-8: the world's largest third-generation synchrotron radiation facility, (3)~Hiroshima University, (4)~RIKEN Fugaku: the Japanese 1st and the world's 4th supercomputer in TOP 500 2024, (5)~Hokkaido University, (6)~Tohoku University, (7)~Science Tokyo, (8)~Nagoya University, (9)~Kyoto University, (10)~Osaka University, and (11)~Kyushu University.

\section{Implementation} \label{sec:infra}
In this section, we describe the implementation of the ARIM-mdx data system.
Fig.~\ref{fig:implementation} shows the implementation details.
The system infrastructure consists of a peta-scale data storage system and the mdx cloud system connected to facilities and supercomputers via SINET6 and the data-transfer IoT devices as its core infrastructure.

On top of this infrastructure, the system provides Jupyter-based containerized computing service and Nextcloud, an open-source cloud storage service. 
The software packages in the Jupyter environment are based on MateriApps LIVE!~\cite{motoyama2022materiapps}, an all-in-one open software suite for materials science.
The middleware components for the elastic container management are implemented based on Kubernetes. 

The key technological significance in the system infrastructure lies in its cloud internal direct connection and IoT-based networking.
\subsection{Cloud Internal Direct Connection}
The primary system-level challenge in this architecture is access latency between the mdx cloud, where the Jupyter environment is deployed for interactive data analysis, and the peta-scale dedicated data storage.
Since data on the dedicated storage are interactively accessed from the Jupyter on the mdx side, very low latency accesses between them are necessary for usability.
In a conventional networking approach, however, the connection between the mdx cloud and storage would typically route through an external network gateway. 
This routing would result in inter-cluster latency, even if the mdx cloud and storage systems are physically located in proximity to each other.

To solve the issue, the ARIM-mdx data system establishes \textit{a cloud internal direct connection}, which creates a direct, dedicated physical network between the mdx cloud and the dedicated storage.
By utilizing layer-2 network virtualization of mdx, we extend the virtual cluster on the mdx to one of its physical ports.
Then, by connecting this port to the dedicated storage, we establish the virtualized intra-network without any impact on the other virtual clusters in the mdx cloud.
The key idea is similar to AWS Direct Connect, Google Dedicated Network, and Azure ExpressRoute, but differs in that it connects directly to the mdx's internal network by bypassing the external network gateway.

\subsection{IoT-based Networking of Standalone Material Facilities}
Another critical system issue is due to the difficulty in establishing the standard network environment for facility operation computers, as discussed in~\ref{sec:secure}.

In the ARIM-mdx data system, we have widely installed \textit{data transfer IoT devices}~\cite{hanai2023cloud} for the direct data transfer from such standalone facility computers.
The key idea behind the networking is to utilize USB volume interface of the standalone facility computers.
The IoT device mimics the USB flash drive to the facility side (using USB OTG \texttt{g\_mass\_gadget}) while the devices are remotely connected to the cloud storage.
Thus, from user's side on facility operation computers, the IoT devices are recognized as USB flash memory, where users just copy their experimental data to the fake USB volume folder.
Once the data are copied to this fake USB volume folder, the IoT devices automatically transfer data to the cloud storage of the system.
Furthermore, by integrating this USB-based data transfer with user authentication, the system can remotely and safely distribute the experimental data to the storage in per-user manner, without any (Ethernet-based) internet connection in the facility computers.

\section{System Evaluation and Statistic} \label{sec:eval}

\subsection{Evaluation for Key Components}
The purpose of this evaluation is to demonstrate the performance improvements achieved by the cloud internal direct connection and the performance characteristics of wide-area networking between the storage, the mdx cloud, and supercomputers. 
For other general performance characteristics for each component, we recommend that readers refer to, for example, the distributed container orchestration~\cite{ferreira2019performance}, SINET6~\cite{kurimoto2023nationwide}, and the data-transfer IoT device~\cite{hanai2023cloud}.

The summary of this evaluation is as follows:
\begin{itemize}
\item The system achieves very low access latency (less than 1ms) from Jupyter service to the storage system, which is 4.7 times faster than conventional approaches.
\item The network connection from the experimental facilities and supercomputers achieves 4.0 times higher throughput than public cloud services.
\end{itemize}

The mdx cloud system includes 100 Gb/s networks for both the external and internal connections. 
SINET6 is built on a 400Gb/s network infrastructure.
The data storage system runs Nextcloud 29.0 powered by Red Hat Enterprise Linux release 8.5 with
CPU Intel(R) Xeon(R) Gold 6348 $\times$ 2 (28 cores $\times$ 2 Sockets) and 64 GB Memory.
Approximately 3 petabytes DDN EXAScaler Storage (Lustre) is installed.
The physical network connection between the mdx cloud and the storage system is based on 10GbE network switches.

\begin{figure}[t]
    \centering
    \begin{minipage}{.47\columnwidth}
          \centering
           \includegraphics[width=\columnwidth]{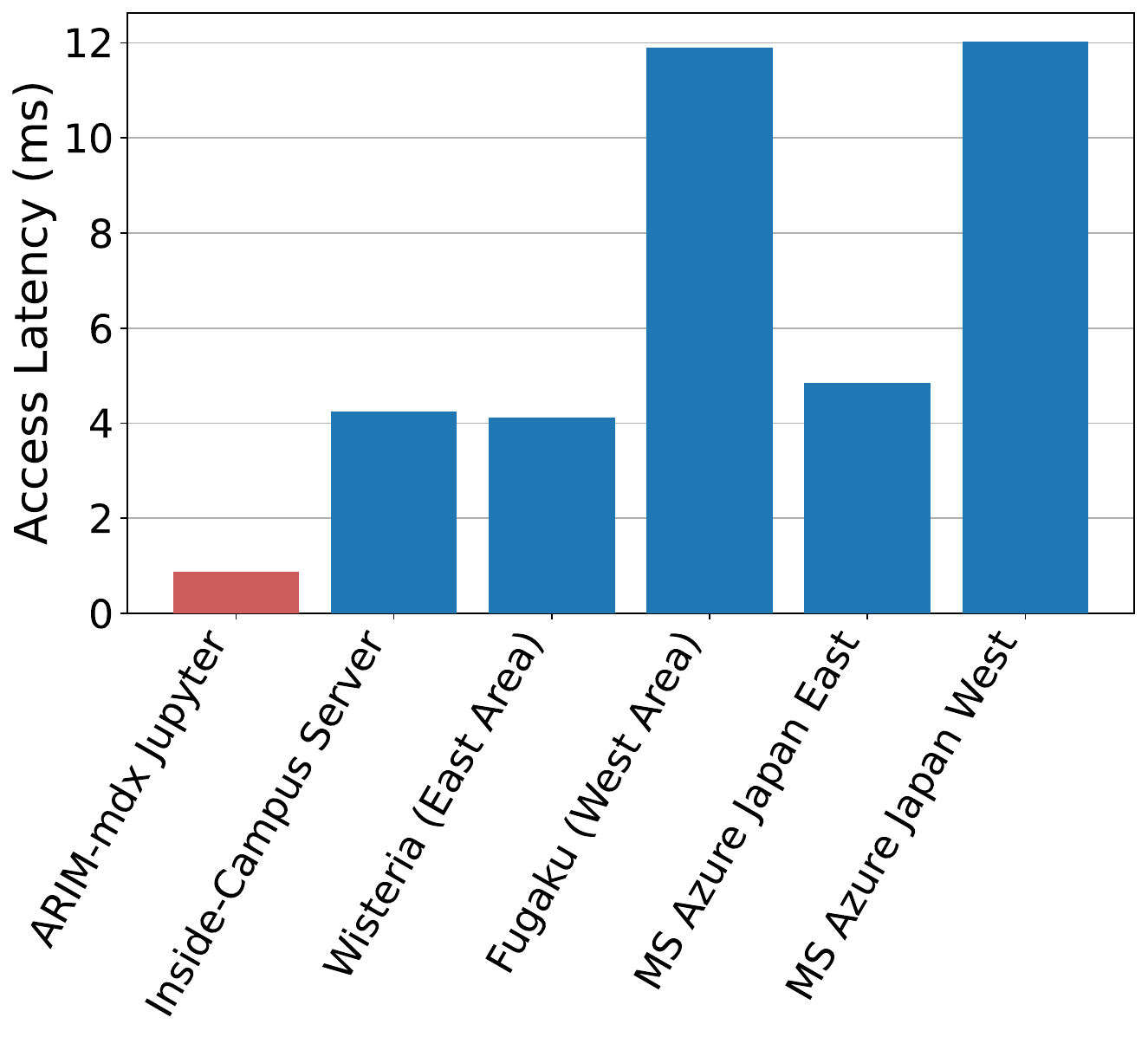}
        \vspace*{\vspacebeforecaption}
        \vspace*{-3pt}
          \caption{Latency to the storage.}
          \label{fig:latency}
    \end{minipage}\hfill
    \begin{minipage}{.51\columnwidth}
       \centering
       \includegraphics[width=\columnwidth]{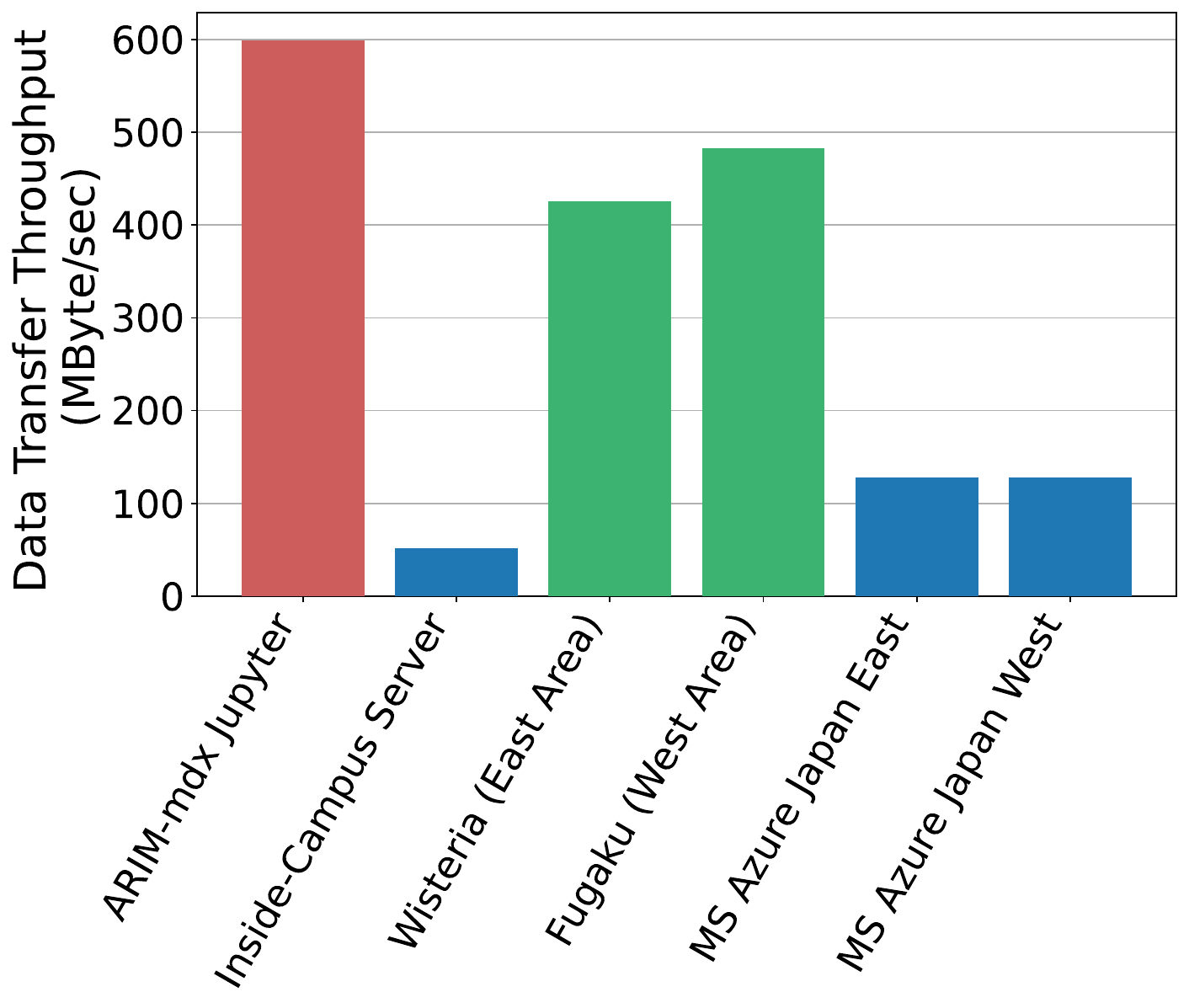}
       \vspace*{\vspacebeforecaption}
       \vspace*{-3pt}
       \caption{Throughput to the storage.}
       \label{fig:throughput}
    \end{minipage}\hfill
    \vspace*{\vspaceaftercaption}
\end{figure}

\begin{figure*}[ht]
    \centering
    \begin{minipage}{0.3\textwidth}
        \centering
        \includegraphics[width=\linewidth]{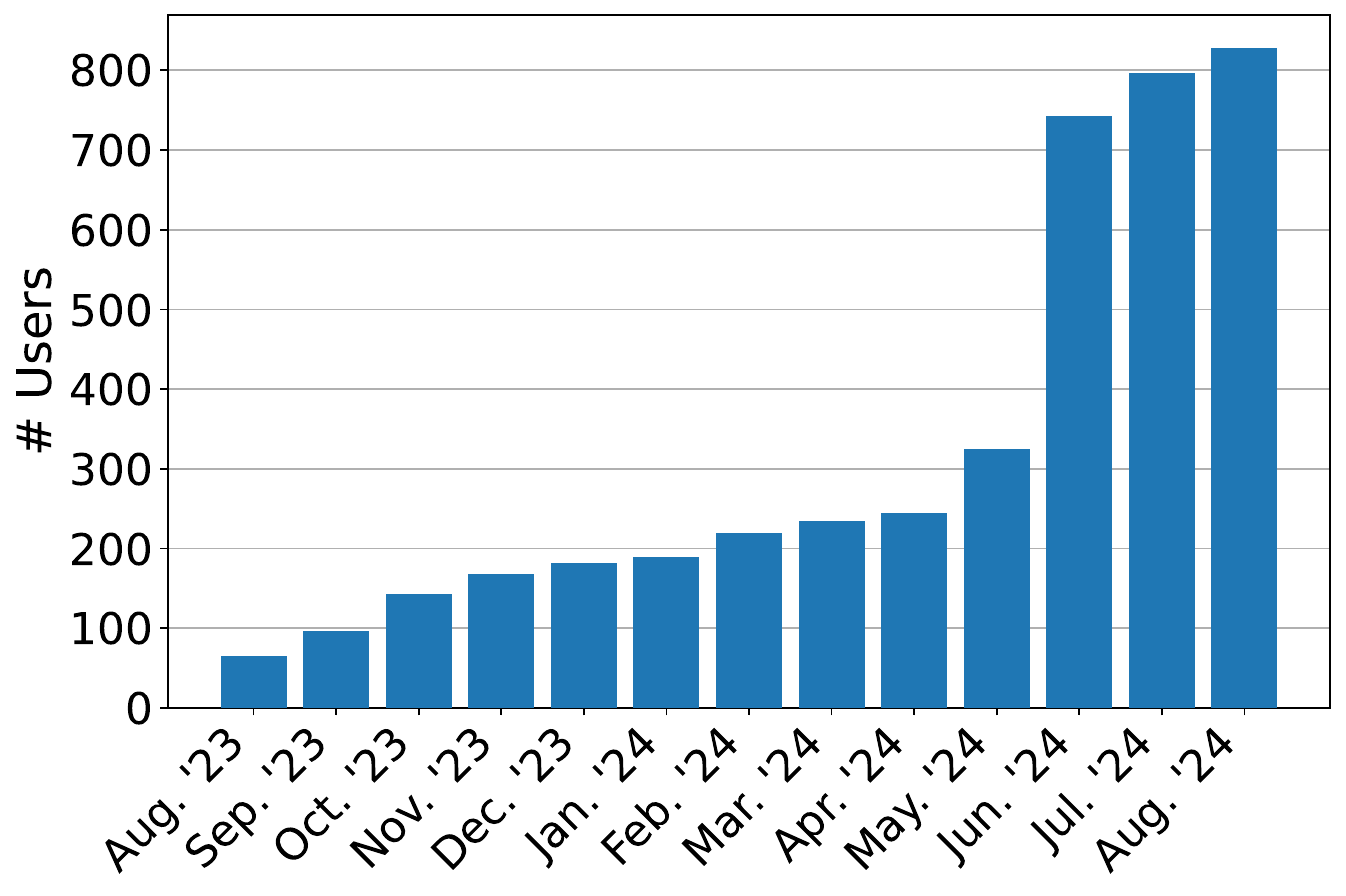}
        \vspace*{\vspacebeforecaption}
        \caption{Number of users.}
        \label{fig:numuser}
    \end{minipage}\hfill
    \begin{minipage}{0.3\textwidth}
        \centering
        \includegraphics[width=\linewidth]{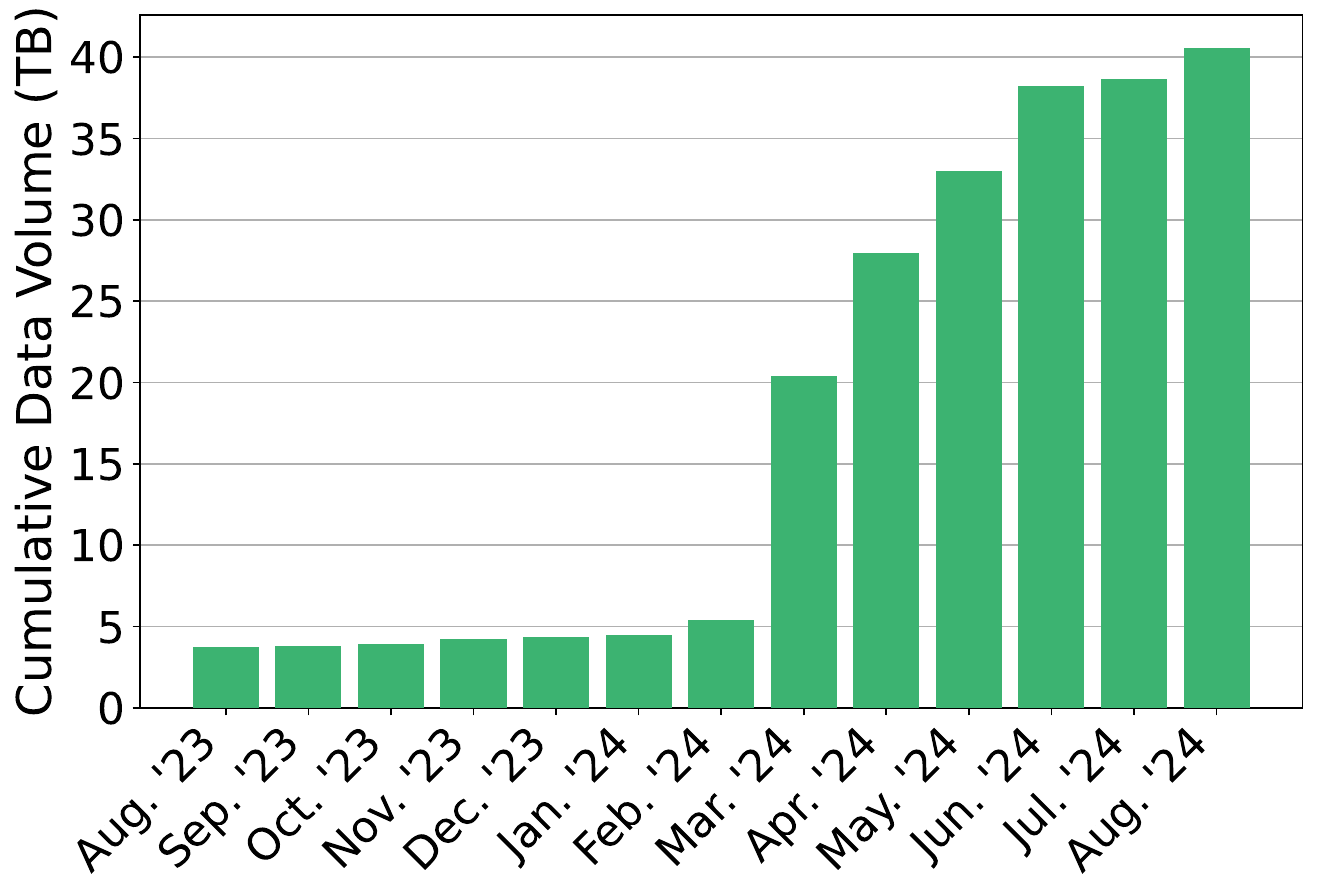}
        \vspace*{\vspacebeforecaption}
        \caption{Data volume.}
        \label{fig:datavolume}
    \end{minipage}\hfill
    \begin{minipage}{0.3\textwidth}
        \centering
        \includegraphics[width=\linewidth]{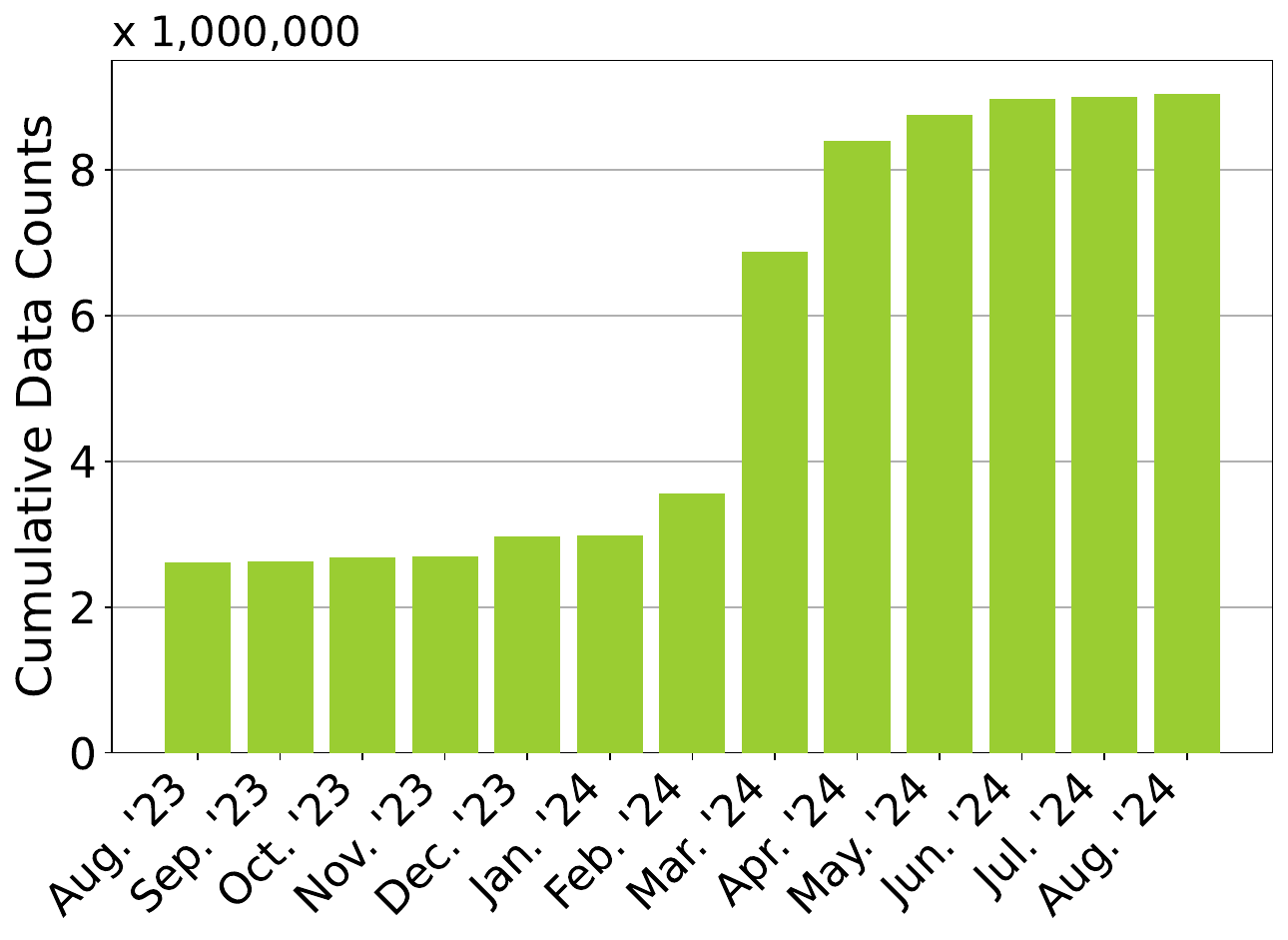}
        \vspace*{\vspacebeforecaption}
        \caption{Number of data files.}
        \label{fig:numfiles}
    \end{minipage}
  \vspace*{\vspaceaftercaption}
\end{figure*}

\smallskip
Fig.~\ref{fig:latency} shows the \texttt{ping} access latency from various sources to the storage system of ARIM-mdx data system.
We conducted the test five times and report the median value.
The variation between tests was less than 5\%.
\texttt{ARIM-mdx Jupyter} shows the access latency from the Jupyter environment on the mdx cloud to the dedicated storage system. Its value is \texttt{0.87} milliseconds (ms). 
Thanks to our established cloud internal direct connection, its access latency is similar performance to the intra-cluster environment (e.g., typically \texttt{0.3}--\texttt{0.5} ms).
\texttt{Inside-Campus Server} shows the access latency from the server, which is physically located in the same building as the mdx cloud (but the network is not directly connected).
Its value is \texttt{4.24} ms due to the fact that the network is through the general route to the mdx cloud.
The packet is once routed to the regional upstream and then goes down to the mdx cloud through the external network gateway.
\texttt{Wisteria (East Area)} shows the access latency from Wisteria supercomputer in the University of Tokyo. 
Its value is \texttt{4.13} ms due to the fact that Wisteria is located in the same building as the mdx cloud.
\texttt{Fugaku (West Area)} shows the latency from Fugaku supercomputer, which is located in Kobe, the western area of Japan. 
Due to the physical distance between Kobe and Tokyo, its value becomes \texttt{11.9} ms.
\texttt{MS Azure Japan East} and \texttt{MS Azure Japan West} show the latency from virtual machines on the Japan East region and Japan West region in Microsoft Azure service.
As the storage server is located in the Tokyo area (the east side of Japan), the latencies become \texttt{4.85} ms for Japan East and \texttt{12.03} ms for Japan West, respectively. 

The overall results reveal that even though a server is physically close to the storage, the access latency is insufficient for interactive computing.
The cloud internal direct connection achieves the intra-cluster level of the access latency, providing the proper usability for the interactive computing.

\smallskip
Fig.~\ref{fig:throughput} shows the data transfer throughput between each server and the storage system of the ARIM-mdx data system.
We transferred 10 files of 1GB each (10GB total) five times using \texttt{Rclone copy} (ver. 1.67 with 4 parallel transfers) and report the median value, with errors less than 5\%.
\texttt{ARIM-mdx Jupyter} shows the throughput between Jupyter container and the storage system on ARIM-mdx data system. 
Thanks to the cloud direct connection, its throughput shows the best performance.
Its value is \texttt{598.8} megabyte per second (MByte per sec).
\texttt{Inside-Campus Server} shows the worst performance due to the bottleneck at its commodity network switches connecting to the campus network.
Even though the intra-campus network uses high-performance infrastructure, the throughput remains low. 
Its value is \texttt{51.65} MByte per sec.
\texttt{Wisteria (East Area)} shows the performance between Wisteria supercomputer and ARIM-mdx data system (storage server). 
The connection from Wisteria to ARIM-mdx data system uses high-performance infrastructure including SINET6.
Thus, it achieves high-throughput and its value is \texttt{425.5} MByte per sec.
Similar to Wisteria, \texttt{Fugaku (West Area)} also achieves very high throughput due to the SINET6.
Its value is \texttt{512.8} MByte per sec.
The slight performance difference between Wisteria and Fugaku may be due to the specification of their storage systems and login nodes where the program was executed.
\texttt{MS Azure Japan East} and \texttt{MS Azure Japan West} show the throughput from the VM in Microsoft Azure, where we use \texttt{Standard\_D4\_v3} instance type (4 vCPUs and 16 GB memory). 
Their values are \texttt{128.3} and \texttt{128.0} MByte per sec, respectively.
Their lower throughput is due to the public Internet and the bandwidth restriction imposed by the cloud provider.

The overall results show the significant benefit of the cloud internal direct connection and SINET6.
By bypassing the public Internet and fully utilizing SINET6, the system is configured to handle large-scale data transfers very efficiently.

\subsection{General Statistic}
We present the statistical data from the one-year operation (from Aug. 2023 to Aug. 2024).
Table~\ref{tab:statistic} shows the overall information for the annual usage.
The number of activated users is 837 from over 140 organizations, where 73.7 \% of the users are from academia while 26.3 \% are from industry.
There are 377 research projects utilizing the ARIM-mdx data system, with the average of approximately 2.48 researchers per project.
The total data volume is approximately 40.6 TB, where 11\% of them are experimental data while 89\% are theoretical data. 
The total number of data files is approximately 9.05 million, with more experimental data files than theoretical ones, in contrast to the data volume distribution.

\begin{table}[ht]
\centering
\caption{Statistic for one-year trial use.}
\vspace*{-8pt}
\begin{tabular}{l r}
\toprule
\# of Users & 839 \\
\ -- \# of Academic Users & 629 \\
\ -- \# of Industrial Users & 210 \\
\# of Research Projects & 377 \\
\# of Organizations for Users & 140+ \\
\midrule
Total Volume of Materials Data & 40.6 TB \\
\ -- Volume of Experimental Data& 4.5 TB \\
\ -- Volume of Theoretical Data& 36.1 TB \\ 
Total \# of Data Files & 9.05 M\\
\ -- \# of Experimental Data Files& 5.82 M \\	
\ -- \# of Theoretical Data Files& 3.23 M \\	
\bottomrule
\end{tabular}
\label{tab:statistic}
\end{table}

Fig.~\ref{fig:numuser}--\ref{fig:numfiles} shows the increase in the total number of users, the volume of data, and the number of data files in the system, respectively.
The number of users increased by \texttt{12.7} times over the course of a year (from Aug. 2023 to Aug. 2024).
The number increases dramatically in May because we intensively conducted tutorial sessions.
The significant increases in the data volume and file numbers in April are due to intensive physical simulations in one theoretical research project.

The overall results show that we successfully attracted a significant number of users during the one-year trial period.
In particular, we successfully garnered significant interest from not only researchers in academic organizations but also industrial organizations.

\begin{figure*}[t]
    \centering
    \begin{minipage}{.485\textwidth}
        \centering
        \includegraphics[width=\linewidth]{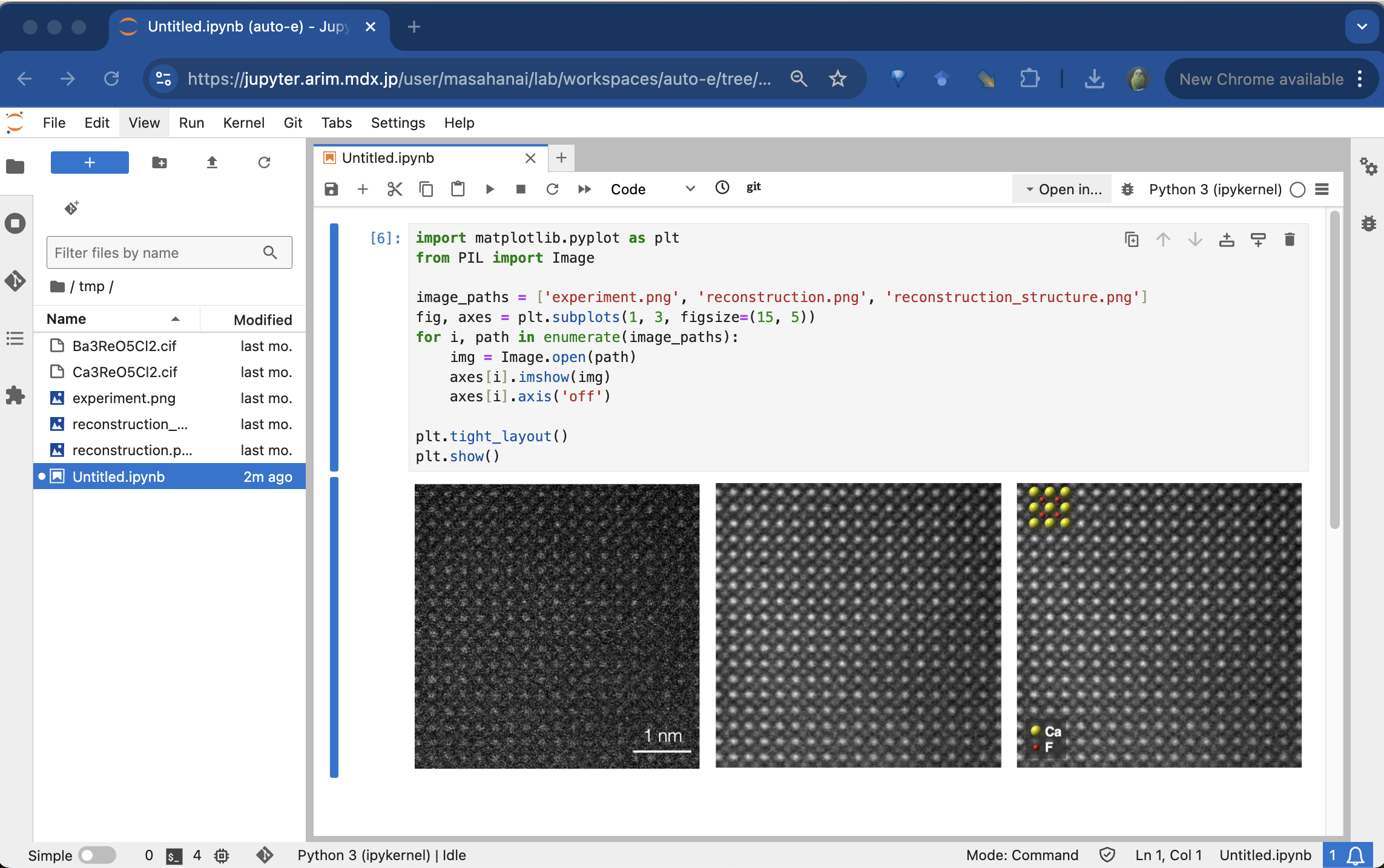}
        \caption{Jupyter Lab of ARIM-mdx data system.}
        \label{fig:jupyter}
    \end{minipage}\hfill
    \begin{minipage}{.485\textwidth}
        \centering
        \includegraphics[width=\linewidth]{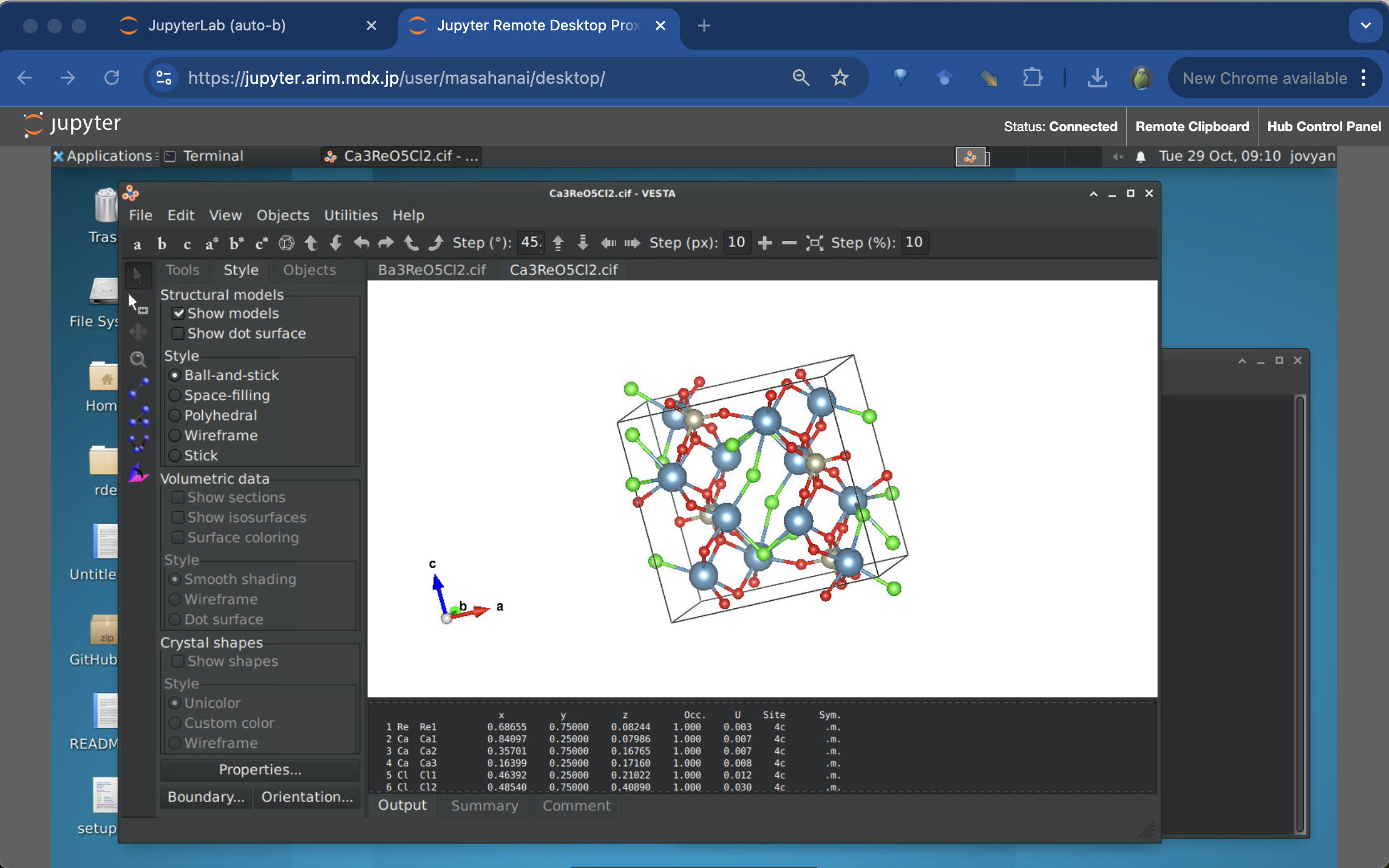}
        \caption{MateriApps LIVE! on Jupyter Remote Desktop.}
        \label{fig:vesta}
    \end{minipage}\hfill
    \vspace*{\vspaceaftercaption}
\end{figure*}

\section{Use Cases} \label{sec:usecase}
In this section, we present several use cases. As these are ongoing, we describe only their key ideas, usage, and system aspects, without details.

\subsection{Denoising Atomic Resolution Image}
\begin{figure}[ht]
  \centering
   \includegraphics[width=.8\columnwidth]{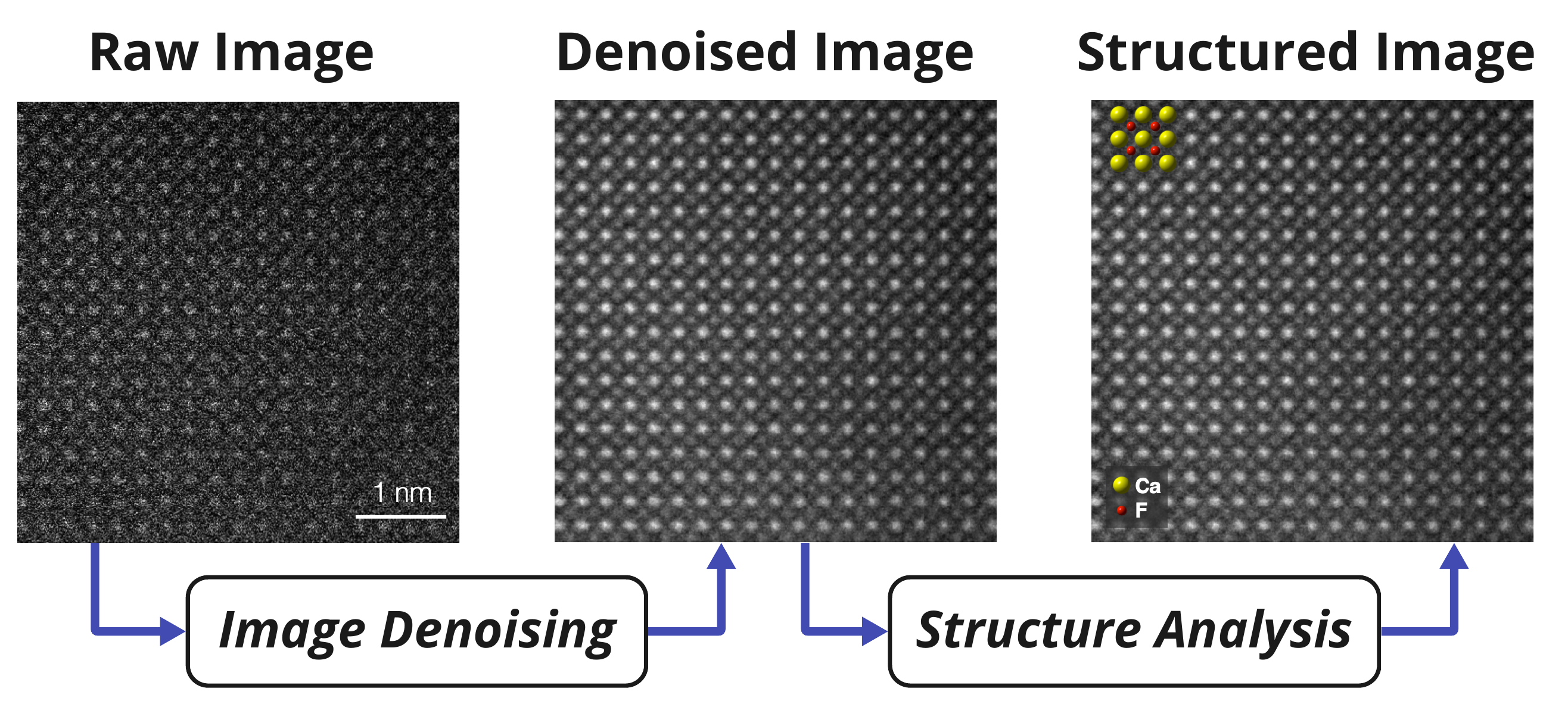}
  \vspace*{-10pt}
  \caption{Denoising for electron microscope image. The configuration of this image is as follows: ADF-STEM CaF2 [001]; 512 $\times$ 512 pixels; accelerating voltage = 200 kV; convergence angle $\alpha$ = 22 mrad~\cite{kawahara2022atomic}.}
  \label{fig:usecase}
\end{figure}

One of the primary goals of atomic resolution electron microscopy is to enhance resolution, and the image denoising is a crucial method~\cite{kusumi2023fast,kusumi2024new} as shown in Fig.~\ref{fig:usecase}.
Raw 2D image signals often contain information unrecognizable to the human eye due to inherent noise.
By processing these raw images to a level where their embedded information becomes interpretable, denoising plays a vital role in supporting experiments, such as identifying optimal observation parts or parameters, and serves as a foundation for post-experimental theoretical analysis of physical phenomena.

Research tasks in image denoising include both the interactive and batch workloads.
For the interactive analysis, the denoising is executed for the target image data during the experiments as shown in Fig.~\ref{fig:jupyter}. 
The image data from the facility computer is transferred via the IoT device. 
On the other hand, the batch processing is conducted on supercomputers for the post-experiment data analysis, where the physical simulation is executed for building the theoretical model.

\subsection{Large-scale Physical Simulations}
The large-scale theoretical data generation and collection for a physical simulation, such as DFT or MD, have been conducted on ARIM-mdx data system and supercomputers.
The typical workflow of the data generation and collection includes the batch workloads on supercomputers and the interactive processing for the post-generation analysis.
As shown in Fig.~\ref{fig:vesta} (using VESTA~\cite{Momma:db5098}), the data collected on the ARIM-mdx data system can be interactively analyzed using GUI tools.
MateriApps LIVE! installed on the Jupyter services includes most of the common tools (simulators, analyzers, and visualizers) for theoretical materials science, so that the developed programs on the ARIM-mdx data system can be immediately deployed and scaled up to supercomputers.

The first specific example of the theoretical data generation is on DFT calculations for the properties of inorganic materials.
In this use case, the large-scale DFT calculations have been conducting on multiple supercomputers in Japan.
We have totally utilized approximately 2M node$\cdot$hours on multiple supercomputers so far.
The execution time per material is approximately from less than half an hour to up to a month, where their wide variation is mainly due to its time complexity $O(n^4)$ for the structure size and variations in their convergence speed.
Small differences in the structure size may cause large difference in the execution time.
In total, 20,000 materials data sets with 5 TB are currently generated.

The second example is that we have been conducting the large-scale MD simulations for electrolytes, which can be utilized in research on Lithium-metal batteries and other related applications~\cite{ko2022electrode}.
In this use case, we computed the MD simulations on mdx, where each node has Intel Xeon Platinum 8368 Processor (38core, 2.4GHz) $\times$ 2 with 256GB memory.
We have totally utilized 12K nodes$\cdot$hours to generate around 650 compositions.
The calculation per composition consumes around 18.4 nodes$\cdot$hours, and the raw data size per compositions is around 42GB.
Currently, around 25.3TB (with meta-data and analysis results) has been generated.

The last concrete example is on the rhenium oxyhalides family, which has exotic optical magnetic properties ($A_3$ReO$_5X_2$, where $A$=Ca, Ba, Sr, Pb, and $X$ = Cl and Br)~\cite{gen2023rhenium}.
The DFT simulation results containing ground-state electronic structure, dielectric function, and effective inter-spin interaction are accumulated and published for later investigation. 
Its computational cost is 3.1K node$\cdot$hours of Wisteria supercomputer in the University of Tokyo and the total amount of the generated data is around 202 GB.

\section{Lessons and Future Directions} \label{sec:future}
Over the course of a year of operation, we have obtained several insights and lessons as follows:

\smallskip
\noindent\textit{\textbf{Safe and efficient data extraction and storing from shared-use experimental facilities are very basic yet most crucial for all experimental researchers:}}
According to the user's survey, over 50\% of users are only interested in the safe and efficient extraction of data, as typically each researcher already has the familiar tools for the data analysis.
The access logs show that the number of daily active users in Nextcloud is 10 times higher than in Jupyter services.

Based on these insights, our future work will focus on developing a high-performance storage system that combines cloud storage interfaces with ultra-fast data handling capabilities, such as Lustre.
This system will prioritize efficient data extraction while maintaining the ease of use found in current cloud storage solutions.
The system will also ensure long-term data integrity and sustainable accessibility in line with the FAIR principles (Findable, Accessible, Interoperable, and Reusable)~\cite{wilkinson2016fair}.
We will enhance security by strengthening encryption and access controls, ensuring compliance with institutional and legal standards.

\smallskip
\noindent\textit{\textbf{Common and standard remote services for experimental materials research are still limited:}}
The comparison between theoretical and experimental researchers reveals a significant disparity in collaborative opportunities.
While theoretical researchers benefit from shared remote tools like supercomputers, bringing active community engagement through system tutorials or workshops, experimental researchers face geographical constraints due to physically separated facilities and limited common remote tools.
This restriction often confines their cross-organizational research exchanges only to general events, such as academic conferences. 

The ARIM-mdx data system has been built as a pioneering nationwide remote tool for experimental researchers, holding great potential to foster and invigorate an extensive research community, similar to what has been achieved with supercomputers in theoretical fields. 
This development could bridge the existing gap in collaborative capabilities between theoretical and experimental researchers, potentially leading to more integrated and diverse research ecosystems in the future.

\smallskip
\noindent\textit{\textbf{Experimental data are inevitably much more sparse and smaller scale than theoretical data:}}
Materials science experiments are inherently costly compared to computational methods and often require manual operations.
The fact leads researchers to focus primarily on areas closely related to their target materials.
Consequently, datasets generated by individual research projects tend to be localized within the broader landscape of materials.
This situation presents a considerable challenge in the pursuit of developing more comprehensive and generalizable machine learning models or databases.
Specifically, to promote cross-material and general data utilization and publishing, addressing data sparsity has become an urgent priority.

For our future work, the ARIM-mdx data system plans to adopt multiple approaches. 
First, we aim to strengthen nationwide data collection efforts to build a more extensive materials database.
Simultaneously, we will pursue integration with high-throughput \textit{automated} experimental systems, enabling efficient and large-scale data generation. 
Furthermore, we plan to complement sparse data through the fusion of open data, academic literature, and large language models (LLMs).
Through these approaches, we aim to create a more comprehensive and versatile database.

\section{Related Work} \label{sec:relatedwork}
Global efforts to promote open data have led to the development of several data repositories for publishing scientific data~\cite{zenodo,dryad,figshare,osf}.
The key difference is that our system focuses primarily on the collection and management of the raw data, rather than publishing the cleaned data.
The ARIM-mdx data system provides a data platform to efficiently prepare the structured dataset for these public repositories.

The integration of large-scale materials research facilities and world-class supercomputers has been undertaken at several national research institutes such as Oak Ridge National Laboratory, Argonne National Laboratory and RIKEN~\cite{salim2019balsam,enders2020cross,antypas2021enabling,bear2021share,veseli2023streaming,al2023cross,brace2023linking,da2024frontiers}.
The key difference is that the ARIM-mdx data system focuses on the nationwide cross-institutional deployment.

Nationwide and cross-institutional service operations for scientific research are well-established in high-performance computing fields.
The key examples are ACCESS~\cite{access} (formerly, XSEDE~\cite{towns2014xsede}) in the United States, EOSC~\cite{budroni2019architectures} in the European Union, and JHPCN~\cite{jhpcn} in Japan.

\section{Concluding Remarks} \label{sec:conclusion}
In this paper, we introduced the ARIM-mdx data system, a nationwide data platform for materials science in Japan.
The ARIM-mdx data system is based on a dedicated high-performance storage system that is tightly integrated with supercomputers, the mdx cloud system, and experimental facilities across Japan via SINET6 and the data-transfer IoT devices. 
The ARIM-mdx data system successfully stores, manages, and analyzes their data, facilitating more effective cross-organizational/domain research collaborations.
As a result, the number of academic and industrial users approaches over 800 during the one-year trial operations.
This highlights its potential as a reference design for large-scale modern materials-science data platforms.

\section*{Acknowledgment}
The authors would like to thank all the contributors\footnote{Contributors \url{https://arim.mdx.jp/en/#devteam}}.
We used mdx~\cite{mdx-IEEECBDCom2022} and the HPCI/JHPCN supercomputers\footnote{Supercomputers \url{https://arim.mdx.jp/en/#supercomputer}}.
This work was supported by MEXT ARIM Japan, MEXT ``Data Creation and Utilisation Type Materials Research and Development Project (DxMT)'' (JPMXP1121467561), ``Joint Usage/Research Center for Interdisciplinary Large-scale Information Infrastructures (JHPCN)'' (jh241010),  MEXT ``Developing a Research Data Ecosystem for the Promotion of Data-Driven Science'' (2024-10) and JSPS KAKENHI (22K17899). 
We used ChatGPT 4o and Claude 3.5 Sonnet for grammar checks and writing revisions.

\bibliographystyle{IEEEtran}
\bibliography{IEEEabrv,ref}

\end{document}

%% file: packages.tex
\usepackage{xcolor}

\usepackage{graphicx}

\usepackage{cite}
\usepackage{booktabs} 
\usepackage{url}

\newcommand{\vspacebeforecaption}{-15pt}
\newcommand{\vspaceaftercaption}{-15pt}

\newcommand{\vspacebeforecaptionsmall}{-7pt}

\usepackage{flushend}

%% file: authorieee.tex
\makeatletter
\newcommand{\linebreakand}{%
  \end{@IEEEauthorhalign}
  \hfill\mbox{}\par
  \mbox{}\hfill\begin{@IEEEauthorhalign}
}
\makeatother

\author{
    \IEEEauthorblockN{
        Masatoshi Hanai\IEEEauthorrefmark{1},
        Ryo Ishikawa\IEEEauthorrefmark{2},
        Mitsuaki Kawamura\IEEEauthorrefmark{1}\IEEEauthorrefmark{8},
        Masato Ohnishi\IEEEauthorrefmark{4},
        Norio Takenaka\IEEEauthorrefmark{7},
    }
     \IEEEauthorblockN{
        Kou Nakamura\IEEEauthorrefmark{7},
        Daiju Matsumura\IEEEauthorrefmark{5},
        Seiji Fujikawa\IEEEauthorrefmark{5},
        Hiroki Sakamoto\IEEEauthorrefmark{6},
        Yukinori Ochiai\IEEEauthorrefmark{3},
    }
    \IEEEauthorblockN{
        Tetsuo Okane\IEEEauthorrefmark{5},
        Shin-Ichiro Kuroki\IEEEauthorrefmark{6},
        Atsuo Yamada\IEEEauthorrefmark{7},
        Toyotaro Suzumura\IEEEauthorrefmark{1},
        Junichiro Shiomi\IEEEauthorrefmark{2},
    }
    \IEEEauthorblockN{
        Kenjiro Taura\IEEEauthorrefmark{1},
        Yoshio Mita\IEEEauthorrefmark{3},
        Naoya Shibata\IEEEauthorrefmark{2},
        Yuichi Ikuhara\IEEEauthorrefmark{2}
   }
 
    \\
    \IEEEauthorblockA{
        \IEEEauthorrefmark{1}
            \textit{Information Technology Center}, 
            \textit{The University of Tokyo}, Tokyo, Japan
    }
    \IEEEauthorblockA{
        \IEEEauthorrefmark{2}
            \textit{Institute of Engineering Innovation}, 
            \textit{The University of Tokyo}, Tokyo, Japan
    }
    \IEEEauthorblockA{
        \IEEEauthorrefmark{3}
            \textit{Department of Electrical Engineering and Information Systems}, 
            \textit{The University of Tokyo}, Tokyo, Japan
    }
    \IEEEauthorblockA{
        \IEEEauthorrefmark{7}
            \textit{Department of Chemical System Engineering},
            \textit{The University of Tokyo}, Tokyo, Japan
    }    
    \IEEEauthorblockA{
        \IEEEauthorrefmark{4}
            \textit{The Institute of Statistical Mathematics},
            \textit{Research Organization of Information and Systems}, Tokyo, Japan
    }
    \IEEEauthorblockA{
        \IEEEauthorrefmark{5}
            \textit{Materials Sciences Research Center}, 
            \textit{Japan Atomic Energy Agency}, Hyogo, Japan
    }
    \IEEEauthorblockA{
        \IEEEauthorrefmark{6}
            \textit{Research Institute for Semiconductor Engineering},
            \textit{Hiroshima University}, Hiroshima, Japan
    }
    \IEEEauthorblockA{
        \IEEEauthorrefmark{8}
            \textit{Center for Emergent Matter Science},
            \textit{RIKEN}, Saitama, Japan
    }
     \IEEEauthorblockA{hanai@ds.itc.u-tokyo.ac.jp}
}
